\documentclass[a4paper,fleqn,usenatbib]{mnras}

\usepackage[T1]{fontenc}
\usepackage{ae,aecompl}
\usepackage{graphicx}
\usepackage{amsmath}
\usepackage{amssymb}
\usepackage{supertabular}

\def\beq{\begin{eqnarray}}
\def\eeq{\end{eqnarray}}

\title[Outflows from black hole hyperaccretion systems]
{Outflows from black hole hyperaccretion systems: short and long-short gamma-ray bursts and ``quasi-supernovae''}

\author[Song et al.]{Cui-Ying Song, Tong Liu\thanks{E-mail:tongliu@xmu.edu.cn}, and Ang Li \\
Department of Astronomy, Xiamen University, Xiamen, Fujian 361005, China}

\date{Accepted XXX. Received YYY; in original form ZZZ}

\pubyear{2018}

\begin{document}
\label{firstpage}
\pagerange{\pageref{firstpage}--\pageref{lastpage}}
\maketitle

\begin{abstract}
The detections of some long gamma-ray bursts (LGRBs) relevant to mergers of neutron star (NS)-NS or black hole (BH)-NS, as well as some short gamma-ray bursts (SGRBs) probably produced by collapsars, muddle the boundary of two categories of gamma-ray bursts (GRBs). In both cases, a plausible candidate of central engine is a BH surrounded by a hyperaccretion disc with strong outflows, launching relativistic jets driven by Blandford-Znajek mechanism. In the framework of compact binary mergers, we test the applicability of the BH hyperaccretion inflow-outflow model on powering observed GRBs. We find that, for a low outflow ratio, $\sim 50\%$, postmerger hyperaccretion processes could power not only all SGRBs but also most of LGRBs. Some LGRBs might do originate from merger events in the BH hyperaccretion scenario, at least on the energy requirement. Moreover, kilonovae might be produced by neutron-rich outflows, and their luminosities and timescales significantly depend on the outflow strengths. GRBs and their associated kilonovae are competitive with each other on the disc mass and total energy budgets. The stronger the outflow, the more similar the characteristics of kilonovae to supernovae (SNe). This kind of `nova' might be called `quasi-SN'.
\end{abstract}


\section{Introduction}\label{sec:intro}

Gamma-ray bursts (GRBs) are classified into two categories divided by $T_{90}\sim 2 ~\rm s$, i.e., long- and short-duration GRBs \citep[LGRBs and SGRBs, see e.g.,][]{Kouveliotou1993}. It is well known that SGRBs are possibly originated from the mergers of neutron star (NS)-NS or black hole (BH)-NS \citep[e.g.,][]{Paczynski1986,Eichler1989,Paczynski1991,Popham1999,Narayan1992,Berger2014}, and LGRBs associated with type Ib/c supernovae (SNe) could be powered by collapsars \citep[e.g.,][]{Tutukov1992,Galama1998,Hjorth2003,Stanek2003,Woosley2006,Campana2006,Fruchter2006,Kumar2015}. With the accumulation of observational data, some LGRBs, such as GRB 060614 \citep[e.g.,][]{Gehrels2006,Zhang2007b,Zhang2009}, were found to be relevant to mergers, while some SGRBs such as GRB 090426 \citep[e.g.,][]{Antonelli2009,Levesque2010,Xin2011} were produced probably by collapsars. Those bursts muddle the LGRB-SGRB boundary. \citet{Zhang2009} summarized the nature of GRBs and proposed a new classification approach mainly based on their progenitors.

Two types of GRB central engines have been widely discussed: hyper-accreting stellar mass BHs \citep[e.g.,][]{Woosley1993,Lei2009,Lei2013,Liu2017a} and rapidly spinning and highly magnetized NSs \citep[magnetars, see e.g.,][]{Usov1992,Thompson1994,Dai1998,Zhang2001,Dai2006}. The detection of gravitational waves (GWs) from close compact binary mergers \citep[e.g.,][]{Eichler1989,Schutz1989,Cutler1994,Lipunov1997,Abadie2010} provides a direct way to verify the progenitors of SGRBs if the association of GWs with SGRBs can be confirmed. After merger events, optical/near-infrared (NIR) emission from the radioactive decay of heavy r-process elements are produced by the merger remains. These transient events are named `kilonovae' \citep{Li1998}. Recently, the NS-NS merger GW event (GW170817) was detected by the advanced LIGO/Virgo collaboration \citep[e.g., ][]{Abbott2017a,Abbott2017b,Alexander2017,Blanchard2017,Coulter2017,Hallinan2017,Troja2017,Shappee2017}. Its electromagnetic counterpart, GRB 170817A accompanied by a kilonova AT 2017gfo, was also discovered \citep[e.g.,][]{Abbott2017c,Chornock2017,Cowperthwaite2017,Evans2017,Kilpatrick2017,Margutti2017,Nicholl2017,Smartt2017}. This event provides the first direct evidence for the progenitor hypothesis of SGRBs.

Outflows may present in accretion processes, especially for super-Eddington accretion discs \citep[e.g.,][]{Shakura1973}.
They were widely studied by theoretical analyses \citep[e.g.,][]{Blandford1999,Liu2008,Gu2015}, numerical simulations \citep[e.g.,][]{Eggum1988,Okuda2002,Ohsuga2005,Ohsuga2011,Jiang2014,Jiang2017,McKinney2014,Sadowski2014,Sadowski2015}, as well as observations \citep[e.g.,][]{Wang2013,Cheung2016,Parker2017a}. \citet{Narayan1994} proposed that the advection-dominated accretion flows (ADAFs) with the positive Bernoulli parameter might generate outflows and, by extension, jets \citep[e.g.,][]{Narayan1995,Abramowicz1995}. \citet{Blandford1999} emphasized the roles of the outflows in ADAFs and developed a variant named advection-dominated inflow-outflow solution (ADIOS). They also constructed 1D and 2D self-similar solutions and found that the structure and radiation of flows were subject to the outflows \citep{Blandford2004}. As to the study of the vertical structures of the discs, strong outflows were required in both optically-thick and optically-thin flows, resulted from energy equilibrium \citep[e.g.,][]{Gu2007,Gu2015}. Outflows can also exist in neutrino-dominated accretion flows \citep[NDAFs, see reviews by][]{Liu2017a}. \citet{Liu2012a} visited the vertical structures and luminosities of NDAFs. They found that outflows might be present in the outer region of the discs, depending on the vertical distributions of the Bernoulli parameter.

A wide variety of mechanisms could lead to the generation of outflows. In optically-thick accretion flows, photons could exert radiation pressure upon materials to blow them away. For the optically-thin cases, such as ADAFs, they might possess a positive Bernoulli parameter due to their high internal energy \citep{Narayan1994,Gu2015}. Moreover, the Blandford-Payne process \citep{Blandford1982} could produce outflows for both optically-thin and optically-thick discs \citep[e.g.,][]{Ma2018}.

Apart from theoretical studies, several simulation results also implied that outflows play essential roles in accretion systems. It was first pointed out by \citet[]{Stone1999}, where 2D hydrodynamic numerical simulations were carried out. Many later simulations confirmed their conclusion, for example in \citet[]{Hawley2001}, \citet{Machida2001}, \citet{Igumenshchev2003}, \citet{Pang2011}, and \citet{Yuan2014}. Furthermore, both the inflow and outflow mass accretion rates decreased inward, following a power-law $\dot{M}\propto r^s$ ($0\leq s\leq 1$). For the outflows, $s\approx 1$ was possible which means that more than $90\%$ of the materials could be pushed into powerful outflows from the accretion disc \citep[e.g.,][]{Yuan2012a,Yuan2012b,Begelman2012}. In the global 3D radiation magneto-hydrodynamical simulation, \citet{Jiang2014} found that the radiation-driven outflows were formed along the rotation axis and about $20\%$ of the radiative energy were carried by outflows. In a recent study of super-Eddington accretion flows onto supermassive BHs (SMBHs), the outflow speed could approach $\sim 0.1-0.4~c$, and the mass flux lost could reach $15\%-50\%$ of the net mass accretion rates \citep{Jiang2017}. The 3D general-relativistic magnetohydrodynamic simulations of NDAFs indicated that the velocities of powerful outflows likely approached $\sim 0.03-0.1~c$ \citep[e.g.,][]{Siegel2017}.

Recent observations also showed the importance of outflows in accretion systems. For Galactic SMBH accretion, more than $99\%$ original gas escaped from the disc \citep{Wang2013}. In quiescent galaxies, \citet{Cheung2016} observed that centrally-driven winds could suppress the star formation. The ultra-fast outflow of the Seyfert I galaxy IRAS 13224-3809 was discovered by \citet{Parker2017a}. They also proposed that the outflows could be detected from the long-term X-ray variability \citep{Parker2017b}. Furthermore, the observed kilonova following GRB 130603B \citep{Berger2013a,Tanvir2013} might also related to disc winds \citep[e.g.,][]{Metzger2014a}.

Outflows play important roles in the BH accretion processes, however, few studies have been done on the BH hyperaccretion system. To investigate the nature of the GRB central engine, we necessarily confront the BH hyperaccretion inflow-outflow model with observational data. In paper I, we constrained the characteristics of the progenitor stars of LGRBs and Ultra-LGRBs with the BH hyperaccretion inflow-outflow model in the collapsar scenario \citep{Liu2018}. In the present work, the applicability of the model is tested to power both SGRBs and LGRBs in the compact binary merger scenario. We further examine the properties of kilonovae triggered by strong neutron-rich outflows. The paper is organized as follows. In Section 2, we describe our BH hyperaccretion inflow-outflow model. The GRB data are presented in Section 3. The properties of kilonovae are shown in Section 4. Section 5 is a brief summary.

\section{Model}

There are two types of GRB central engine candidates widely discussed: Magnetar and hyperaccreting BH with stellar mass. After the mergers of NS-NS or BH-NS, an NS with high spin (period $\sim1~ \rm ms$) and high surface magnetic field ($\sim10^{15}~\rm G$) might be formed, known as a millisecond magnetar. The released spin-down energy could power GRBs. Alternatively, in the BH accretion disc models, the GRB jet can be produced either through the neutrino-antineutrino annihilation process \citep[see e.g.,][]{Popham1999,Di2002,Narayan2001,Liu2007}, or via the electromagnetic processes.

Some general-relativistic magnetohydrodynamics simulations have showed the evidences for the Blandford-Znajek (BZ) mechanism \citep[][]{Blandford1977,Blandford1977a} in GRB central engines \citep[e.g .,][]{Nagataki2009,Tchekhovskoy2012}. \citet{Barkov2010} has confirmed the possibility of the magnetically driven stellar explosions, and has pointed out the required magnetic flux in excess of $10^{28}~\rm G~cm^{2}$ in the close compact binary scenario for LGRBs. Moreover, some studies have shown that the BZ mechanism is more efficient than the neutrino annihilation if the magnetic fields are strong enough or the accretion rates are lower than the ignition accretion rates of NDAFs \citep[e.g.,][]{Kawanaka2013,Liu2015a,Lei2017,Liu2017a}. The BH spin energy might be extracted via the BZ mechanism when a strong magnetic field ($\sim10^{13}-10^{15} ~\rm G$) threads the spinning BH and is connected with a distant astrophysical load. Essentially all central engine models require a strong, large-scale magnetic field to launch GRBs. It may inherit and redistribute the large-scale magnetic field of the merging components following the magnetic flux conservation \citep[e.g.,][]{Liu2016,Punsly2016}.

In the BH hyperaccretion scenario, introducing the effects of the outflows and the relativistic jets driven by the BZ mechanism, the disc model is called BH hyperaccretion inflow-outflow model. Two parameters are required for describing a hyperaccreting stellar mass BH: the dimensionless mass $m_{\rm BH}=M_{\rm BH}/M_{\rm \sun}$ and spin $a_* \equiv cJ_{\rm BH}/GM_{\rm BH}^{2}$.

Then the BZ jet power can be estimated as \citep[e.g .,][]{Lee2000a,Lee2000b,Li2000,McKinney2005,Barkov2008,Komissarov2009,Barkov2010,Lei2013}
\beq
L_{\rm BZ}=1.7\times10^{50}a_{*}^{2}m_{\rm BH}^{2}B_{\rm BH,15}^{2}F(a_*){\rm~erg~s^{-1}},
\eeq
where $B_{\rm BH,15}=B_{\rm BH}/10^{15} {\rm G}$ and $F(a_*)=[(1+q^{2})/q^{2}][(q+1/q)\arctan(q)-1]$ with $q=a_{*}/(1+\sqrt{1-a_{*}^{2}})$.

We can evaluate the magnetic field strength when the magnetic pressure on the BH horizon balances the ram pressure of the innermost part of the disc \citep[e.g.,][]{Moderski1997}
\beq
\frac{B_{\rm BH}^2}{8\pi}=P_{\rm ram}\sim \rho c^2 \sim \frac{\dot{M}_{\rm in}c}{4\pi r_{\rm BH}^2},
\eeq
where $r_{\rm BH}=GM_{BH}(1+\sqrt{1-a_*^{2}})/c^2$ denotes the event horizon of the BH, and $\dot{M}_{\rm in}$ is the accretion rate at the inner boundary. Then the magnetic field strength can be written as
\beq
B_{\rm BH} \simeq 7.4 \times 10^{16}\dot{m}_{\rm in}^{1/2}m_{\rm BH}^{-1}(1+\sqrt{1-a_{*}^{2}})^{-1}~ {\rm G}.
\eeq

Inserting above equation into Equation (1), we obtain the BZ jet power as a function of mass accretion rate at  and spin of BH,
\beq
L_{\rm BZ}=9.3\times10^{53}a_*^{2}\dot{m}_{\rm in} X(a_*){\rm~erg~s^{-1}},
\eeq
and
\beq
X(a_*)=F(a_*)/(1+\sqrt{1-a_*^{2}})^{2}.
\eeq

The dimensionless BH mass accretion rate at the innermost stable orbit $\dot{m}_{\rm in}$ is defined as
\beq
\dot{m}_{\rm in}=\frac{(1-f) m_{\rm disc}}{T_{90,\rm rest}},
\eeq
where $T_{\rm 90,\rm rest} = T_{90}/(1+z)$ is the duration of the prompt emission in the rest frame, and $z$ stands for the redshift. $m_{\rm disc}=M_{\rm disc}/M_{\odot}$ stands for the dimensionless accretion disc mass, and $f$ represents the fraction of the outflow mass to the disc mass. As mentioned in the Introduction, outflows have been found to be very strong in many studies, therefore we take $f = 99\%$ for the strongest outflow case, and $f=50\%$ as a typical value for comparison.

Recently, the simulations of NS-NS mergers \citep[e.g.,][]{Dietrich2015} and BH-NS mergers \citep[e.g.,][]{Foucart2014,Just2015,Kyutoku2015,Kiuchi2015,Siegel2017} showed that the remnant disc mass $m_{\rm disc}$ likely possessed an upper limit $\sim 0.3~M_{\rm\sun}$, which depends on the equation of state of the NS, the mass ratio, the total mass and the period of the binary \citep[e.g.,][]{Oechslin2006,Dietrich2015}. We therefore take $m_{\rm disc}=0.01, ~0.1, ~0.3$ in our calculations.

On the other hand, the jet power can be obtained from the observational data of GRBs \citep[e.g.,][]{Fan2011,Liu2015b,Song2016}, i.e.,
\beq
L_{\rm j}\simeq\frac{(E_{\rm \gamma,\rm iso} + E_{\rm k,\rm iso})(1+z)\theta_{\rm j}^{2}}{2 T_{90}},
\eeq
where $E_{\rm \gamma,\rm iso}$, $E_{\rm k,\rm iso}$, and $\theta_{\rm j}$ denote the isotropic radiated energy, the isotropic kinetic energy of afterglows, and the jet opening angle, respectively.

\begin{figure}
\includegraphics[width=\columnwidth]{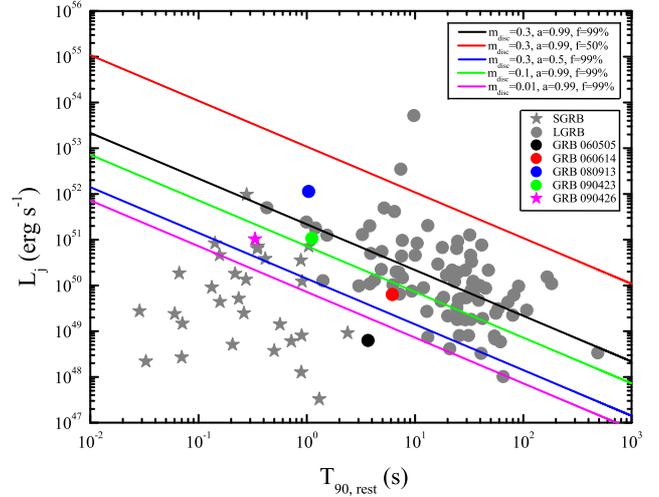}
\caption{Luminosities and timescales of BZ jets originated from compact binary mergers in five solid lines, different colors corresponding to different parameters as labelled. The gray filled stars and circles denote SGRBs and LGRBs data, respectively. The magenta star stands for SGRB that may be related to the collapse of massive stars. The colorful circles are for LGRBs possibly from NS-NS or BH-NS mergers.}
\end{figure}

\section{GRB Data}

As shown in Table 1, we collected the data of $T_{90}$, $z$, $E_{\rm \gamma,\rm iso}$, $E_{\rm k,\rm iso}$, $\theta_{\rm j}$ and the peak energy in the rest frame $E_{\rm p, rest}$ of 30 SGRBs and 89 LGRBs. It is worth noting that the measurements of $E_{\rm \gamma,\rm iso}$, $E_{\rm k,\rm iso}$, and $\theta_{\rm j}$ are model dependent. Considerable debates surround the origins of some GRBs due to their perplexing observational phenomena \citep[e.g.,][]{Zhang2009,Xin2011,Kann2011,Li2016}, some unusual GRBs are labelled by the superscript $*$ in the table, which are shown as following:

\textbf{GRB 060505} This burst has a duration of $4\pm1\rm~s$, the low energy $\sim 10^{49} \rm~erg$ and a very low redshift $z=0.0894$. The presence of a SN is ruled out down to limit of hundreds times fainter than SN 1998bw \citep[]{Ofek2007,Fynbo2006}. The star formation rate, metallicity, ionization state of the host environment are more similar to SGRBs than to LGRBs \citep[]{Levesque2007}.

\textbf{GRB 060614} Its $T_{90}\sim100 \rm~s$ in the BAT (15-150~keV) band groups it with LGRBs, while its peak luminosity and temporal lag completely satisfy the SGRB subclass \citep[]{Gehrels2006}. The low-star-formation-rate host galaxy \citep[]{Savaglio2009}, and its irrelevant to any known SN \citep[]{GalYam2006,Fynbo2006,DellaValle2006} suggest that it is related to a compact binary merger rather than a collapsar \citep[e.g.,][]{Zhang2007a}. Furthermore, the discovery of a NIR bump in afterglow denotes the strong connection between GRB 060614 and a kilonova, and provides tangible evidence to support the merger origin \citep[]{Yang2015,Jin2015,Horesh2016}.

\textbf{GRB 080913} The burst duration $T_{90}$ in the BAT band is $8\pm1\rm~s$. Considering its high redshift $z=6.7$, the rest-frame duration of this burst is $T_{90, \rm rest}\sim1 \rm~s$. \citet{Palshin2008} fitted the Konus-Wind and the \textit{Swift}/BAT joint spectral analysis, and derived that the best fit peak energies are $E_{\rm peak}=131_{-48}^{+225} \rm~keV$ and $E_{\rm peak}=121_{-39}^{+232} \rm~keV$ for the cutoff power law and Band-function spectra, respectively. Placing this GRB at $z=1$, it can be classified as SGRBs due to intrinsically short duration and hard spectrum. \citet{PerezRamirez2010} presented the X-ray, NIR and millimetre observations, and proposed that the progenitor of this burst was likely from a BH-NS merger. A maximally-rotating BH might form in the center and power this GRB by the BZ mechanism. However, this GRB is consistent with the lag-luminosity correlation and the Amati relation of LGRBs, which makes the collapsar origin cannot be ruled out \citep[]{Greiner2009,Zhang2009}.

\textbf{GRB 090423} Similar to GRB 080913, this burst was measured with a high redshift $z=8.6$ and a BAT band duration $T_{90}=10.3\pm1.1 \rm~s$. In rest frame, the peak energy and duration are $491\pm200 \rm~keV$ \citep{Amati2009} and $\sim1.1 \rm~s$, respectively.

\textbf{GRB 090426} It is a SGRB with an observed duration of $T_{90}\sim1.28 \rm~s$ at $z=2.609$ \citep{Antonelli2009}. On the other hand, the soft spectrum $E_{\rm p, rest}=177_{-65}^{+90} \rm~keV$ \citep{Amati2009} and  burst environment are similar to those of LGRBs. The number density of the medium ($>11.2 \rm~cm^{-3}$) is not consistent with the condition of the compact-binary-merger progenitors, which often occur in low density medium \citep[e.g.,][]{Xin2011}.

Above all, the observed LGRBs and SGRBs are contaminated by each other. Some studies also found that the duration distributions of SGRBs and LGRBs overlaped each other \citep[e.g.,][]{Horvath2002}. Certainly $T_{90}$ is not a good criterion to manifest the nature of GRBs, which resulted in the old dichotomy between the LGRBs related to collapsars and the SGRBs originated from compact binary mergers.

Figure 1 shows the luminosities and timescales of the BZ jets originated from the compact binary mergers. The solid lines in different colors correspond the BZ jets with the different values of $m_{\rm disc}$, $a_{*}$, and $f$. The gray filled stars and circles denote SGRBs and LGRBs data, respectively. The magenta star stands for the SGRB which may related to the collapse of massive stars. The colorful circles are for LGRBs possibly from NS-NS or BH-NS mergers. By comparing the red and black lines, one can find that the luminosities of SGRBs decrease when the outflow increases. For $f = 50\%$, most of LGRBs find their place under our predicted lines of the BZ jets. For $f = 99\%$, at least half of those LGRBs cannot be explained by the model. Therefore, our model can explain not only all SGRBs but also most of LGRBs (with a low outflow ratio).

Actually, assuming that $T_{90}$ is near or proportional to the duration of the central engine activity, then the duration of GRBs may be closely related to the properties of progenitors. The collapsar scenario is suggested to produce LGRBs through accretion because of the typical envelope fallback timescale is $10\rm~s$ \citep[e.g.,][]{MacFadyen1999}. The BH accretion disc systems after NS-NS/BH-NS mergers have a typical accretion timescale $\sim 0.01-0.1 \rm~s$, which were raised to account for SGRBs in many models \citep[e.g.,][]{Narayan2001,Aloy2005}. However, the discover of X-ray flares \citep[e.g.,][]{Burrows2005,Nousek2006,Wu2007,Chincarini2010,Margutti2010,Mu2016a,Mu2016b}, extended emission \citep[e.g.,][]{Lazzati2001,Connaughton2002,Norris2010,Norris2011,Liu2012b} and plateaus \citep[e.g.,][]{Troja2007,Rosswog2007,Rowlinson2013} in some GRBs denote that the duration of the GRB central engine activity is much longer than $T_{90}$ in both LGRBs and SGRBs. The gamma-ray duration $T_{90}$ may much shorter than the central engine activity duration, named `tip-of-iceberg' effect \citep{Lv2014,Zhang2014,Li2016,Gao2017a,Liu2018}. A longer accretion timescale is needed to explain these observations by the BH hyperaccretion systems. The durations of the compact binary mergers are not necessarily to be ``short", and in principle the collapsar model can also bring forth SGRBs \citep[e.g.,][]{Janiuk2008,Zhang2009}. If it is the case, the values of the jet luminosities and timescales of GRBs are larger than these in Table 1. Thus the admission of the model limitations is stricter than in the current situations.

\begin{table*}
\centering
\caption{GRBs data}
{\footnotesize
\def\arraystretch{1.0}
\tabcolsep=6.6pt
\begin{tabular}[width=1.0 \linewidth]{ccccccccc}
\hline
\hline
Name & $T_{90}$ & $z$ & $E_{\rm \gamma,\rm iso}$ & $E_{\rm k,\rm iso}$ & $E_{\rm p,\rm rest}$ & $\theta_{\rm j}$ & $L_{\rm j}$ &  Ref  \\
\hline
  & (s) & z & ($10^{52}$\rm erg) & ($10^{52}$\rm erg) &(\rm keV) & (rad) & ($10^{50} \rm erg\rm~s^{-1}$) &  \\
\hline
SGRBs   \\
\hline
050509B  &	0.04 &	0.225 &	$0.00024_{-0.0001}^{+0.00044}$ &	0.0055 &	$100.45_{-98}^{+748.475}$  &	$ >0.05 $ &	0.022 &	1, 2 \\	 																																																																				050709  &	0.07 &	0.161 &	$0.0027\pm0.0011$ &	0.0016 &	$96.363_{-13.932}^{+20.898}$ &	$ >0.26 $ &	0.2397 &	1, 3 \\																																																																					050724A &	3 &	0.257 &	$0.009_{-0.002}^{+0.011}$ &	0.027  &	$138.27_{-56.565}^{+502.8}$ &	$ >0.35 $ &	0.0915 &	1, 2 \\																																																																					051210 &	1.3 &	1.3  &	$0.4_{-0.2}^{+0.5}$  &	0.238 &	$943_{-598}^{+1495}$ &	$>0.05$ &	0.1411 &	1, 2 \\																																																																		051221A &	1.4 &	0.5465 &	$0.28_{-0.1}^{+0.21}$ &	1.26  &	$603.135_{-293.835}^{+1020.69}$ &	0.12 &	1.2234 &	1, 2 \\
060502B &	0.09 &	0.287 &	$0.003_{-0.002}^{+0.005}$ &	0.012 &	$437.58_{-244.53}^{+926.64}$ &	$>0.05$ &	0.0268 &	1, 2 \\																																																																					060801 &	0.5 &	1.13 &	$0.7_{-0.5}^{+1.5}$ &	0.071 &	$1320.6_{-724.2}^{+2279.1}$ &	$0.0561_{-0.0063}^{+0.0056}$ &	0.5167 &	1, 2, 4 \\ 																																																																					061006 &	0.4 &	0.438 &	$3_{-1}^{+4}$ &	0.314 &	$819.66_{-402.64}^{+1308.58}$ &	$0.407_{-0.173}^{+0.068}$ &	97.3211 &	1, 2, 4 \\ 																																																																					061201 &	0.8 &	0.111 &	$3_{-2}^{+4}$ &	0.007 &	$666.6_{-388.85}^{+888.8}$ &	0.017 &	0.0603 &	1, 2 \\																																																																					061210 &	0.2 &	0.409 &	$0.09_{-0.05}^{+0.16}$ &	0.086 &	$760.86_{-436.79}^{+1070.84}$ &	$>0.37$ &	8.3909 &	1, 2, 5 \\																																																																					070429B &	0.5 &	0.902 &	$0.07_{-0.02}^{+0.11}$ &	0.451 &	$228.24_{-125.532}^{+1418.89}$ &	$>0.05$ &	0.2477 &	1, 2 \\																																																																					070714B &	2.0 &	0.923 &	$1.16_{-0.22}^{+0.41}$ &	0.232 &	$2153.76_{-730.74}^{+1499.94}$ &	$0.33_{-0.11}^{+0.11}$ &	7.2217 &	 1, 3, 4 \\ 																																																																					070724A &	0.4 &	0.457 &	$0.003\pm0.001$ &	0.099 &	$~99$ &	$0.27_{-0.16}^{+0.16}$ &	1.346 &	 1, 3, 4 \\ 																																																																					070729 &	0.9 &	0.8  &	0.017 &	0.132 &	$840.6_{-351}^{+1526}$ &	$>0.05$ &	0.0372 &	1, 6 \\																																																																					070809 &	1.3 &	0.473  &	0.0056 &	0.391 &	$75.6_{-13.9}^{+12.6}$ &	$0.4_{-0.333}^{+0.08}$ &	3.5473 &	 1, 4, 6 \\ 																																																																					071227 &	1.8 &	0.381  &	$0.1\pm 0.02$ &	0.025 &	$1384 \pm 277$ &	$>0.0262$ &	0.0033 &	 1, 5, 7  \\ 																																																																					080905A &	1.0 &	0.122  &	0.0005 &	0.0024 &	$502.8_{-280.5}^{+950.6}$ &	$0.28_{-0.16}^{+0.15}$ &	0.0127 &	 1, 4, 6 \\ 																																																																					$090426^{*}$ &	1.2 &	2.609 &	$0.5\pm0.1$ &	13.5 &	$177 _{-65}^{+90}$ &	0.07 &	10.3115 &	1, 8 \\																																																																					090510 &	0.3 &	0.903  &	$4.47_{-3.77}^{+4.06}$ &	0.307 &	$7490_{-494.8}^{++532.8}$ &	0.017 &	0.4379 &	1, 6 \\																																																																					090515 &	0.04 &	0.403  &	0.0008 &	0.062 &	$90.1_{-16.8}^{+47.4}$ &	$>0.05$ &	0.2753 &	1, 6 \\																																																																					100117A &	0.3 &	0.92  &	$0.09 \pm 0.01$ &	0.11 &	$551_{-96}^{+142}$ &	$0.27_{-0.15}^{+0.15}$ &	4.6373 &	1, 4, 7  \\																																																																					100206A &	0.1 &	0.408  &	$0.0763_{-0.0229}^{+0.0789}$ &	0.0073 &	$638.98_{-131.21}^{+131.21}$ &	$>0.05$ &	0.1471 &	1, 9 \\																																																																					100625A &	0.3 &	0.453  &	$0.075\pm0.003$ &	0.0093 &	$701.32 \pm 114.71$ &	$>0.05$ &	0.051 &	1, 10 \\																																																																					101219A &	0.6 &	0.718  &	$0.49 \pm 0.07$ &	0.045 &	$842_{-136}^{+177}$ &	$0.29_{-0.14}^{+0.14}$  &	6.3966 &	1, 4, 7  \\																																																																					111117A &	0.5 &	1.3 &	$0.338 \pm 0.106$ &	0.377 &	$966 \pm 322$ &	0.105 &	1.8114 &	1, 10 \\																																																																					120804A &	0.81 &	1.3 &	0.7 &	0.7 &	$310.5_{-66.7}^{+151.8}$ &	$>0.19$ &	7.1539 &	11 \\																																																																					130603B   &	0.18 &	0.356 &	$0.212 \pm 0.023$ &	0.28 &	$900 \pm 140$ &	0.07 &	0.9077 &	1,  10 \\																																																																					131001A &	1.54 &	0.717 &	0.037 &	0.541 &	$94.44\pm 24.04$ &	$>0.05$ &	0.0805 &	1, 12 \\																																																																					140622A &	0.13 &	0.959 &	0.0065 &	0.977 &	$86.2\pm 15.67$ &	$>0.05$ &	1.8522 &	1, 13 \\																																																																					160821B &	0.48 &	0.16 &	$0.021\pm0.002$ &	$~8$ &	$97.44\pm 22.04 $ &	$~0.063$ &	3.8455 &	14, 15 \\																																																																																																																																																																																													\hline																																																																			\end{tabular}
}

\end{table*}

 \begin{table*}
 \contcaption{}
 \label{tab:continued1}
\begin{tabular}[width=1.0 \linewidth]{ccccccccc}
\hline
\hline
Name & $T_{90}$ & $z$ & $E_{\rm \gamma,\rm iso}$ & $E_{\rm k,\rm iso}$ & $E_{\rm p,\rm rest}$ & $\theta_{\rm j}$ & $L_{\rm j}$ &  Ref  \\
\hline
      & (s) & z & ($10^{52}$\rm erg) & ($10^{52}$\rm erg) &(\rm keV) & (rad) & ($\rm 10^{50} erg\rm~s^{-1}$) &  \\
\hline
 LGRBs  \\
\hline
 970508 &	$14.0 \pm 3.6$ &	0.8349 &	$0.61 \pm 0.13$ &	$0.99\pm0.14$ &	$145 \pm 43 $ &	$0.3775\pm0.0291$ &	1.4765 &	16, 17 \\																																																																					970828 &	160.0 &	0.96 &	$29 \pm 3$ &	    37.154 &	$586 \pm 117$ &	0.1239 &	0.6212 &	16, 18 \\																																																																					971214 &	$31.23\pm 1.18$ &	3.418 &	$21 \pm 3 $ &	$8.48\pm 0.97$ &	$685 \pm 133$ &	$>0.0967\pm 0.0040$ &	1.9483 &	16, 17 \\																																																																					980613 &	$42.0 \pm 22.1$ &	1.0964  &	$0.59 \pm 0.09 $ &	 $1.22\pm0.38$ &	$194\pm89$ &	$ >0.2194\pm 0.0101$ &	0.2166 &	16, 17 \\																																																																					980703 &	$76.0 \pm 10.2$ &	0.966 &	$7.2 \pm 0.7 $ &	 $2.41\pm0.63$ &	$503 \pm 64$ &	$0.1957\pm 0.0141$ &	0.4745 &	16, 17 \\																																																																					990123 &	$63.3 \pm 0.3$ &	1.61 &	$229 \pm 37 $ &	534 &	$1724 \pm 446$ &	$0.064 \pm 0.005$ &	6.4408 &	16, 19 \\																																																																					990510 &	$67.58 \pm1.86$ &	1.619 &	$17 \pm 3 $ &	$13.16\pm1.12$ &	$423 \pm 42$ &	$0.0586\pm 0.0037$ &	0.2006 &	16, 17 \\																																																																					990705 &	$32.0 \pm 1.4$ &	0.84 &	$18 \pm 3 $ &	 $0.34\pm0.12$ &	$459 \pm 139$ &	$0.0930\pm 0.0072$ &	0.4557 &	16, 17 \\																																																																					991216 &	$15.17\pm0.091$ &	1.02  &	$67 \pm 7 $ &	$ 36.64\pm1.79$ &	$648\pm134$ &	 $0.0798\pm0.0126$ &	4.3918 &	16, 17 \\																																																																					000210 &	$9.0 \pm1.4$ &	0.846 &	$14.9 \pm 1.6 $ &	 $0.50 \pm 0.12$ &	$753 \pm 26$ &	$> 0.1194\pm0.0049$ &	2.2489 &	16, 17 \\																																																																					000926 &	$1.30 \pm 0.59$ &	2.0387 &	$27.1\pm5.9 $ &	$9.97 \pm 3.75$ &	$310\pm20$ &	$0.1075\pm0.0054$ &	50.0191 &	16, 17 \\																																																																					010222 &	$74.0 \pm4.1$ &	1.4769  &	$81 \pm 9 $ &	 $22.79\pm2.48$ &	$766 \pm 30$ &	$0.0559\pm0.0023$ &	0.5426 &	16, 17 \\																																																																					011211 &	$51.0 \pm7.6$ &	2.14 &	$5.4\pm0.6 $ &	 $71.32\pm0.22$ &	$186\pm24$ &	$0.1114\pm0.0070$ &	2.9279 &	16, 17 \\																																																																					020813 &	89 &	1.25 &	$66 \pm 16 $ &	   $204.174$ &	$590 \pm 151$ &	0.0541 &	0.9993 &	16, 18 \\																																																																					021004 &	$77.1 \pm2.6$ &	2.3304  &	$3.3 \pm 0.4 $ &	$8.35\pm 1.45$ &	$266 \pm 117$ &	$0.2211\pm0.0787$ &	1.225 &	16, 17 \\																																																																					050126 &	30 &	1.29 &	$0.8_{-0.2}^{+1.0} $ &	$39.8\pm80.4$ &	$387.01_{-144.27}^{+1135.84}$ &	$0.365_{-0.125}^{+0.095}$ &	20.4159 &	2, 4, 20 \\																																																																					050315 &	$96 \pm10$ &	1.9500  &	$5.7_{-0.1}^{+6.2} $ & $512.403_{-65.577}^{+45.299}$ &	$126.85_{-123.9}^{+32.45}$ &$0.0759_{-0.0091}^{+0.0080}$ &	4.5837 &	2, 17 \\																																																																					050318 &	$32\pm2$ &	1.4436  &	$2.2 \pm 0.16 $ &	$11.259_{-0.685}^{+0.867}$ &	$115\pm25$ &	$0.0380\pm0.0070$ &	0.0742 &	16, 17 \\																																																																					050319 &$139.4 \pm 8.2$ &3.2425 &$4.6_{-0.6}^{+6.5}$ &	$77.896_{-28.695}^{+20.496}$ &$190.912_{-182.428}^{+114.548}$ &$0.0380_{-0.0070}^{+0.0051}$ &	0.1812 &	2, 17 \\																																																																					050401 &	38 &	2.9 &	$35 \pm 7$ &	    $4570.9\pm1317.6$ &	$467\pm110$ &	$0.472_{-0.044}^{+0.02}$ &	5168.5851 &	4, 16, 20  \\																																																																					050416A &	5.4 &	0.654 &	$0.1 \pm 0.01$ &	    $15.1\pm 3.9$ &	$25.1\pm4.2$ &	$0.237_{-0.059}^{+0.114}$ &	13.0142 &	4, 16, 20  \\																																																																					050505 &	$63\pm2$ &	4.27 &	$16_{-3}^{+13}$ &$237.829_{-49.203}^{+98.405}$ &	$737.8_{-226.61}^{+1807.61}$ &	$0.0290_{-0.0030}^{+0.0059}$ &	0.8928 &2, 17 \\																																																																					050525  &	11.5 &	0.606 &	$2.5 \pm 0.43$ &	    $28.2\pm 8.1$ &	$127\pm10$ &	$0.0551_{-0.0062}^{+0.0069}$ &	0.6507 &	4, 16, 20  \\																																																																					050730 &	$155\pm20$ &	3.97 &	$9_{-3}^{+8}$ &	86.1223 &	$974.12_{-432.39}^{+2798.11}$ &	$>0.023$ &	0.0807 &	2, 19 \\																																																																					050802 &	20 &	1.71 &	$1.8197_{-0.30614}^{+1.6477}$ &	    $616.6\pm 295.4$ &	$268.3_{-75.9}^{+623.3}$ &	$0.29_{-0.15}^{+0.15}$ &349.8991 &	4, 20,21 \\																																																																					050814 &	48 &	5.3 &	$6_{-1}^{+3}$ &	831.764 &	$403.2_{-138.6}^{+378}$ &	0.0419 &	9.6506 &	2, 18 \\																																																																					050820A &	600 &	2.615 &	$97.4 \pm 7.8$ &	53.7145 &	$1325\pm277$ &	0.184 &	1.5369 &	16, 19 \\																																																																					050904 &	$183.6\pm13.2$ &	6.295  &	$124 \pm 13$ &	$88.37_{-44.2}^{+86.3}$ &	$3178 \pm 1094$ &	$0.0340\pm0.0051$ &	0.4877 &	16, 17 \\																																																																					050922C &	4.5 &	2.198 &	$5.3\pm1.7$ &	47.725 &	$415\pm111$ &	0.026 &	1.2736 &	16, 19 \\																																																																					051109A &	360 &	2.35 &	$6.5 \pm 0.7$ &	   169.824 &	$539\pm200$ &	0.0593 &	0.2884 &	16, 18 \\																																																																					060124 &	$298 \pm2$ &	2.297  &	$41 \pm 6$ &	 $578.87_{-12.66}^{+110.79}$ &	$784\pm285$ &	$0.0531_{-0.0040}^{+0.0091}$ &	0.9666 &	16, 17 \\																																																																					060206 &	$5.0\pm0.7$ &	4.05 &	$4.3 \pm 0.9$ &	$386.76 \pm 93.02$ &	$394 \pm 46$ &	$0.0351\pm 0.0010$ &	24.3279 &	16, 17 \\																																																																					060210 &	$220 \pm 70$ &	3.91 &	$42_{-8}^{+35}$ &	1313.2261 &	$667.76_{-191.49}^{+1703.77}$ &	$0.024 \pm 0.002$ &	0.871 &	2, 19 \\																																																																					060418 &	$52 \pm 1$ &	1.49 &	$13 \pm 3$ &	7.5307 &	$572 \pm 143$ &	$0.029 \pm 0.006$ &	0.0413 &	16, 19 \\		
 $060505^{*}$&	$4\pm1$	&0.089 &	$0.0012\pm0.0002$ &	0.028&	$482.4_{-167.7}^{+524.8}$ &	$\sim0.4$&	0.06275 & 6, 22, 23\\		 																																																																																																																																					060526 &$258.8\pm5.4$ &	3.21 &	$2.6 \pm 0.3$ &	$15.58_{-0.21}^{+0.24}$ &	$105 \pm 21$ &	$0.0630\pm0.0010$ &	0.0587 &	16, 17 \\																																																																					060605 &	$19 \pm 1$ &	3.8 &	$2.5_{-0.6}^{+3.1}$ &	115 &	$681.6_{-240}^{+1723.2}$ &	$>0.046$ &	3.14 &	2, 19 \\																																																																					060607A &	$100 \pm 5$ &	3.082 &	$9_{-2}^{+7}$ &	0.822 &	$567.398_{-167.362}^{+889.876}$ &	$>0.095$ &	0.1808 &	2, 19 \\																																																																					$060614^{*}$ &	6.9 &	0.12 &	$0.21 \pm 0.09$ &	1.698 &	$55 \pm 45$ &	0.2025 &	0.6328 &	16, 18 \\																																																																					060707 &	210.0 &	3.42 &	$5.4 \pm 1$ &	   102.329 &	$279 \pm 28$ &	$0.1379$ &	2.1525 &	16, 18 \\																																																																					060714 &	$108.2\pm6.4$ &	2.71 &	$7.7_{-0.9}^{+7.5}$ &	$250.46 \pm 248.11$ &	$196.63_{-181.79}^{+348.74}$ &	$0.0201\pm0.0010$ &	0.1788 &	2, 17 \\																																																																					060908 &	$18.0\pm0.8$ &	1.8836 &	$9.8 \pm 0.9$ &	$2017.68_{-504.42}^{+2522.09}$ &	$514 \pm 102$ &	$0.0080_{-0.0010}^{+0.0051}$ &	1.0394 & 16, 17 \\																																																																					061007 &	$75\pm5$ &	1.262 &	$86 \pm 9$ &	29.9425 &	$890 \pm 124$ &	$>0.138$ &	3.3244 &	16, 19 \\																																																																					061021 &	79.0 &	0.35 &	$10_{-4}^{+8}$ &	    6.166 &	$661.5_{-337.5}^{+985.5}$ &	0.1501 &	0.3106 &	2, 18 \\																																																																					061121 &	$81\pm 5$ &	1.314 &	$22.5 \pm 2.6$ &	20.5215 &	$1289 \pm 153$ &	0.099 &	0.6018 &	16, 19 \\																																																																					061222A &	16.0 &	2.09 &	$21_{-4}^{+11}$ &	2290.868 &	$710.7_{-210.12}^{+747.78}$ &	0.0471 &	49.5146 &	2, 18 \\
 070110 &	$89 \pm 7$ &	2.352 &	$3.0_{-0.5}^{+2.5}$ &	0.687 &	$372.072_{-90.504}^{+1035.77}$ &	$>0.274$ &	0.518 &	2, 19 \\																																																																					070125 &	$63.0\pm1.7$ &	1.5477 &	$80.2 \pm 8$ &	$6.43_{0.17 }^{0.9}$ &	$934 \pm 148$ &	$0.2304\pm0.0105$ &	9.2574 &	16, 17 \\																																																																					070306 &	3.0 &	1.5 &	$6_{-1}^{+5}$ &	67.608 &	$300_{-97.5}^{+1340}$ &	0.0768 &	18.081 &	2, 18 \\																																																																					070318 &	$63\pm3$ &	0.84 &	$0.9_{-0.2}^{+0.9}$ &	47.2719 &	$360.64_{-143.52}^{+818.8}$ &	$0.127 \pm 0.008$ &	1.1331 &	2, 19 \\																																																																					070411 &	$101\pm5$ &	2.95 &	$10_{-2}^{+8}$ &	83.6596 &	$474_{-154.05}^{+2196.2}$ &	$0.032 \pm 0.005$ &	0.1875 &	2, 19 \\																																																																					070508 &	23.4 &	0.82 &	$8_{-1}^{+2}$ &	    10.715 &	$378.56_{-74.62}^{+138.32}$ &	0.0611 &	0.2716 &	2, 18 \\																																																																					071010A &	$6\pm1$ &	0.98 &	$0.13\pm0.01$ &	7.2164 &	$73_{-69.9}^{+97.7}$ &	$0.090 \pm 0.008$ &	0.9812 &	19, 21 \\																																																																					071010B &	$35.7\pm0.5$ &	0.947 &	$1.7 \pm 0.9$ &	7.2713 &	$101 \pm 20$ &	$0.150 \pm 0.006$ &	0.5494 &	16, 19 \\																																																																					071031 &	150.5 &	2.692 &	$3.9\pm0.6$ &	1.554 &	$45.23_{-41.54}^{+22.85}$ &	$0.070 \pm 0.013$ &	0.0328 &	19, 21 \\		
 \hline																																																																																																																																								\end{tabular}
\end{table*}

 \begin{table*}
 \contcaption{}
 \label{tab:continued2}
\begin{tabular}[width=1.0 \linewidth]{ccccccccc}
\hline
\hline
Name & $T_{90}$ & $z$ & $E_{\rm \gamma,\rm iso}$ & $E_{\rm k,\rm iso}$ & $E_{\rm p,\rm rest}$ & $\theta_{\rm j}$ & $L_{\rm j}$ &  Ref  \\
\hline
 & (s) & z & ($10^{52}$\rm erg) & ($10^{52}$\rm erg) &(\rm keV) & (rad) & ($10^{50} \rm erg\rm~s^{-1}$) &  \\
\hline																																																																																																																																							080310 &	32.0 &	2.43 &	   6.0256 &	29.512 &	$75.4_{-30.8}^{+72}$ &	0.0628 &	0.7509 &	18, 21\\																																																																					080319C &	29.55 &	1.95 &	$14.1 \pm 2.8$ &	74.4078 &	$906 \pm 272$ &	$>0.102$ &	4.5924 &	16, 19 \\																																																																					080330 &	$61 \pm 9$ &	1.51 &	$0.21\pm0.05$ &	21.0923 &	$<88$ &	$>0.087$ &	0.3315 &	19, 24 \\																																																																					080413B &	8.0 &	1.1 &	$2.4 \pm 0.3$ &	138.038 &	$150 \pm 30$ &	0.1047 &	20.1874 &	18, 25 \\																																																																					080603A &	150 &	1.688 &	$2.2\pm0.8$ &	52.5129 &	$160_{-130}^{+920}$ &	$0.071 \pm 0.011$ &	0.247 &	19, 26\\																																																																					080710 &	$120\pm17$ &	0.845 &	$0.8\pm0.4$ &	2.6451 &	200 &	$>0.062$ &	0.0102 &	19, 27 \\																																																																					080810 &	$108 \pm5$ &	3.35 &	$45 \pm 5$ &	41.8519 &	$1470 \pm 180$ &	$>0.105$ &	1.9266 &	19, 25 \\																																																																					$080913^{*}$ &	$8\pm1$ &	6.695 &	$8.6\pm2.5$ &	$\leqslant 10$ &	$710\pm350$ &	$0.359_{+0.099}^{-0.125}$ &	114.0568 &	4, 25 \\																																																																					081008 &	$162.2\pm25.0$ &	1.967 &	$9.98_{-2.31}^{+2.34}$ &	$134.7_{-17.3}^{+18.3}$ &	$255.9\pm57.47$ &	$0.0227\pm0.0070$ &	0.0682 &	17, 28 \\																																																																					081203A &	223 &	2.1 &	$35 \pm 3$ &	11.2261 &	$1541 \pm 757$ &	$>0.116$ &	0.4319 &	19, 29 \\																																																																					081222 &	5.8 &	2.77 &	$30 \pm 3$ &	131.826 &	$505 \pm 34$ &	0.0489 &	12.5737 &	18, 25 \\																																																																					090313 &	$78 \pm 19$ &	3.375 &	3.2 &	276.8523 &	$240.1_{-223.5}^{+885.4}$ &	$>0.093$ &	6.7881 &	19, 21 \\																																																																					090323 &	$133.1\pm1.4$ &	3.568  &	$410\pm50$ &	$116_{-9}^{+13}$ & $1901 \pm 343$ &	$0.0489_{-0.0017}^{+0.0070}$ &	2.1579 &	17, 25 \\																																																																					090328 &	$57\pm 3$ &	0.7354 &	$13 \pm 3$ &	$82_{-18}^{+28}$ & $1028 \pm 312$ &	$0.0733_{-0.0140}^{+ 0.0227}$ &	0.7767 &	17, 25 \\																																																																					$090423^{*}$ &	$10.3\pm 1.1$ &	8.23  &$11 \pm 3$ &	$340_{-140}^{+110}$ &$491 \pm 200$ &	$0.0262_{-0.0052}^{+0.0122}$ &	10.7949 &	17, 25 \\																																																																					090424 &	$49.47 \pm 0.9$ &	0.544 &	$4.6 \pm 0.9$ &	53.1215 &	$273 \pm 50$ &	$>0.378$ &	12.718 &	19, 25 \\																																																																					090812 &	75.9 &	2.452 &	$40.3 \pm 4$ &	148.827 &	$2023 \pm 663$ &	$>0.071$ &	2.1671 &	19, 29 \\																																																																					090902B & $19.328 \pm 0.286$ &	1.8829  &	$1.77\pm0.01$ &	$56_{-7}^{+3}$ &$596.76\pm17.2974$ &	$0.0681\pm0.0035$ &	1.9973 &	3, 17 \\																																																																					090926A &	$20\pm 2$ &	2.1062 &	$210_{-8}^{+9}$ &	$6.8 \pm 0.2$ &	$1279.75\pm62.124$ &	$0.1571_{-0.0349}^{+0.0698}$ &	41.4656 &	3, 17 \\																																																																					091018 &	106.5 &	0.97 &	 0.5888 &	    12.023 &	$51.29\pm23.7$ &	0.0820 &	0.0784 &	18, 28  \\																																																																					091020 &	65 &	1.71 &	$12.2 \pm 2.4$ &	    51.286 &	$129.809 \pm 19.241$ &	0.1204 &	1.9162 &	7, 18 \\																																																																					091024 &	1020 &	1.092 &	$28 \pm 3$ &	37.2529 &	$794 \pm 231$ &	$>0.071$ &	0.0337 &	19, 29 \\																																																																					091029 &	39.2 &	2.752 &	$7.4 \pm 0.74$ &	40.303 &	$230 \pm 66$ &	$>0.192$ &	8.39 &	19, 29 \\																																																																					091208B &	71 &	1.063 &	$2.01 \pm 0.07$ &	    50.119 &	$297.4846_{-28.6757}^{+37.13}$ &	0.1274 &	1.2276 &	7, 18 \\																																																																					100418A &	$8.0 \pm 2.0$ &	0.6235 &	$0.99_{-0.34}^{+0.63}$ &	3.36 &	$47.08_{-43.83}^{+3.247} $ &	0.3560 &	5.5352 &	30, 31 \\																																																																					100621A &	$63.6 \pm 1.7$ &	0.542 &	$4.37 \pm 0.5$ &	111.7596 &	$146 \pm 23.1$ &	$>0.234$ &	7.6734 &	19, 29 \\																																																																					100728B &	$12.1 \pm 2.4$ &	2.106 &	$2.66 \pm 0.11$ &	95.665 &	$406.886 \pm 46.59$ &	$>0.063$ &	5.0071 &	7, 19 \\																																																																					100901A &	439 &	1.408 &	6.3 & 167.3233 &	230	&0.152 &	1.098 &	19, 32 \\																																																																					100906A &	$114.4\pm1.6$ &	1.727 &	$28.9 \pm 0.3$ &	23.8173 &	$289.062_{-55.0854}^{+47.72}$ &	$0.055 \pm 0.002$ &	0.19 &	7, 19 \\																																																																					110205A &	$257 \pm 25$ &	2.22 &	$56 \pm 6$ &	31.2172 &	$715 \pm 239$ &	0.064 &	0.2237 &	19, 29 \\																																																																					110213A &	$48 \pm 6$ &	1.46 &	$6.9 \pm 0.2$ &	25.7527 &	$242.064_{-16.974}^{+20.91}$ &	$>0.142$ &	1.6843 &	7, 19 \\																																																																					120119A &	$70 \pm 4$ &	1.728 &	36 &	4.17 &	$498.9 \pm 22.31$ &	$0.032 \pm 0.002$ &	0.0801 &	19, 28 \\																																																																					120326A &  	$11.8 \pm 1.8$ &	1.798 &	$3.2\pm0.1$ &	$14.0 \pm 0.07$ &	$152\pm14$ &	$0.0803\pm0.0035$ &	1.3142 &	17, 33 \\																																																																					120521C  &	$26.7\pm0.4$ &	6 &	$8.25_{-1.96}^{+2.24}$ &	$22_{-14}^{+37}$ & $682_{-207}^{+845}$ &	$0.0524_{-0.0192}^{+0.0401}$ &	1.0885 &	17, 34 \\
\hline																																																																			\end{tabular}
\begin{minipage}{16cm}
\emph{Notes}: \\
$^{\star}$ GRBs have unusual characteristics on observations \citep[]{Xin2011,Zhang2009}.

\emph{References}: \\
(1) \citealt{Liu2015a}; (2) \citealt{Butler2007}; (3) \citealt{Zhang2011}; (4) \citealt{Ryan2015}; (5) \citealt{Racusin2009}; (6) \citealt{Kann2011} ; (7) \citealt{Zhang2012}; (8) \citealt{Antonelli2009}; (9) \citealt{Tsutsui2013}; (10) \citealt{Zaninoni2016}; (11) \citealt{Berger2013b}; (12) \citealt{Cummings2013}; (13) \citealt{Sakamoto2014}; (14) \citealt{Stanbro2016} ;(15) \citealt{Lv2017}; (16) \citealt{Amati2008}; (17) \citealt{Song2016}; (18) \citealt{Nemmen2012}; (19) \citealt{Yi2016}; (20) \citealt{Zhang2007b}; (21) \citealt{Kann2010}; (22) \citealt{Xu2009}; (23) \citealt{Ofek2007}; (24) \citealt{Guidorzi2009}; (25) \citealt{Amati2009}; (26) \citealt{Guidorzi2011}; (27) \citealt{Kruhler2009}; (28) \citealt{Dichiara2016}; (29) \citealt{Ghirlanda2012}; (30) \citealt{Marshall2011}; (31) \citealt{Laskar2015};
  (32) \citealt{Gorbovskoy2012}; (33) \citealt{Demianski2017}; (34) \citealt{Yasuda2017}.\\
\end{minipage}
\end{table*}

\section{Quasi-SNe}

As strong GW sources in the nearby galaxies, the NS-NS/BH-NS mergers are expected to have electromagnetic counterparts such as SGRBs \citep[e.g.,][]{Eichler1989,Nakar2007,Berger2014,Kumar2015,Levan2016}, off-axis emission of SGRB jets \citep[e.g.,][]{Rhoads1999,Lazzati2017,Xiao2017}, optical/NIR signals powered by the decay of heavy radioactive elements in the ejection matter \citep[e.g.,][]{Li1998, Metzger2017}, radio flares \citep[e.g.,][]{Nakar2011,Gao2013,Piran2013}, or X-ray emission from GRB central engine \citep[e.g.,][]{Nakamura2014,Kisaka2015}. Furthermore, \citet[]{Zhang2013} proposed that potentially an early X-ray afterglow would continue for thousands of seconds followed GW bursts once the NS-NS merger produced a magnetar rather than a BH. So the existence of an X-ray transient might be used as a criterion to judge if it is a remnant magnetar. The BH scenario will not fit the data in this case \citep[]{Sun2017}.

As mentioned above, the compact objects merger model is accompanied by the ejection of neutron-rich matter. The dynamical ejecta, in a typical timescale of milliseconds, constitute contact-interface materials which are squeezed out by the hydrodaynamic force \citep[e.g.,][]{Oechslin2007,Bauswein2013,Hotokezaka2013} or the tidal force \citep[e.g.,][]{Kawaguchi2015}. For NS-NS mergers, the typical ejecta velocity and mass are in the range of $\sim 0.1-0.3~c$ and $\sim 10^{-4}-10^{-2}~M_{\rm \sun}$, respectively \citep[e.g.,][]{Hotokezaka2013}. Recent BH-NS merger simulations revealed that the ejecta mass could reach $0.1~M_{\rm \sun}$ with a similar velocity as in the NS-NS cases \citep[]{Kawaguchi2015,Kawaguchi2016}. Then heavy radioactive elements will form via the r-process of neutron-rich matter. The radioactive decay of these elements provides a source for powering transient optical/NIR emission \citep[]{Eichler1989,Li1998}, named `kilonova' \citep[]{Kulkarni2005} [also called `macronova' \citep[]{Metzger2010}].

In addition to the radioactivity of the merger ejecta, the remnant materials' fall-back accretion \citep[e.g.,][]{Rosswog2007,Rossi2009,Chawla2010,Kyutoku2015}, the ejecta from the disc, like winds \citep[e.g.,][]{Metzger2012,Ma2018} or outflows, and magnetars \citep[e.g.,][]{Zhang2013,Gao2013,Yu2013,Metzger2014,Gao2015,Gao2017b,Yi2017,Yi2018} can also power kilonovae. For some SGRBs with extended emission or internal X-ray plateaus, the magnetars might form after NS-NS mergers and provide the additional energy injection into the ejecta to power `mergernovae' \citep[e.g.,][]{Yu2013}. In this paper, the nature of the outflows represents the heavy-nuclei-dominated injections into kilonovae.

The kilonovae are claimed to be detected in the optical/NIR band, associated with some GRBs, i.e., GRBs 050709 \citep[]{Jin2016}, 060614 \citep[]{Yang2015,Jin2015,Horesh2016}, 130603B \citep[]{Tanvir2013,Berger2013a,Fan2013}, and 160821B \citep[]{Kasliwal2017}. The excess optical emission was also discovered in GRB 080503 with a lack of redshift \citep[]{Perley2009}. \citet[]{Gao2017b} revisited the \textit{Swift} SGRB samples and found three `magneter-powered mergernova' candidates, i.e., GRBs 050724, 070714B, and 061006. The luminosities of these sources are ten times or a hundred times higher than those of typical kilonovae. For the recent GW event, the luminosity of GRB 170817A is one order of magnitude lower than that of associated AT 2017gfo \citep[e.g.,][]{Smartt2017}.

The outflow matter is much massive than the dynamical ejecta after mergers, so we just calculated the effects of the outflows on kilonovae. Following \citet[]{Li1998}, we adopt a power law decay model here, and assume the material envelope expanding uniformly with the fixed velocity $V$, constant outflow mass $M_{\rm outflow}=fM_{\rm disc}$, surface radius $R$, and density $\rho$. The critical time $t_{c}$, when the optical depth of the expanding sphere satisfies $\kappa \rho R = 1$, can be calculated as
\beq
t_{\rm c}&=&(\frac{3\kappa fM_{\rm disc}}{4\pi V^{2}})^{1/2} \nonumber
\\&=&1.13~{\rm day}~(\frac{fM_{\rm disc}}{0.01~M_{\odot}})^{1/2}(\frac{3V}{c})^{-1}(\frac{\kappa}{\kappa_{e}})^{1/2},
\eeq
where $\kappa \sim \kappa_{e}$ and $\kappa_{e} \sim 0.1~\rm cm^{2}~ g^{-1}$ \citep[e.g.,][]{Metzger2010} represent the average opacity and the electron scattering, respectively, and $V = 0.1~c$ is adopted. As shown in Figures 1 and 2, we take $m_{\rm disc}$=0.3, 0.1, 0.01, and $f=99\%, ~50\%$ to demonstrate the budget on the outflow strength and remnant inflow mass to kilonovae and GRBs. The luminosity of a kilonova powered by the radioactive decay of nuclei can be estimated as
\beq
L_{\rm kilo}=L_{0}\sqrt{\frac{8\beta}{3}}Y\left (\sqrt{\frac{3}{8\beta}}\tau\right),
\eeq
where $L_{0}=3\eta fM_{\rm disc}c^{2}/(4\beta t_{\rm c})$, $\tau=t/t_{\rm c}$, and $\beta=V/c$. $\eta =3\times 10^{-6}$ denotes the fraction of rest-mass energy released in radioactive decay \citep[]{Metzger2010}, and $Y(x)=e^{-x^2}\int_0^x e^{k^2}dk$ is the Dawson's integral.

By assuming blackbody emission, the effective temperature of the thermal emission is given by
\beq
T_{\rm eff}=\left(\frac{L_{\rm kilo}}{4\pi\sigma R_{\rm ph}^2}\right)^{1/4},
\eeq
where $\sigma$ is the Stephan-Boltzmann constant, and $R_{\rm ph}$ is the photosphere radius corresponding the radius of mass shell when the optical depth is equal to 1.

The flux density of the source at photon frequency $\nu$ can be described as follows,
\beq
F_{\nu}=\frac{2\pi h{\nu}^3}{c^2}\frac{1}{e^{h\nu/kT_{\rm eff}}-1}\frac{R_{\rm ph}^2}{D^2}.
\eeq
Then we can get the luminosity of an observational frequency $\nu$, i.e.,
\beq
\nu L_{\nu}=\nu 4\pi D^2 F_{\rm \nu}=\frac{8{\pi}^2 h{\nu}^4 R_{\rm ph}^2 }{c^2}\frac{1}{e^{h\nu/kT_{\rm eff}}-1}.
\eeq

\begin{figure}
\centering
\includegraphics[scale=0.51]{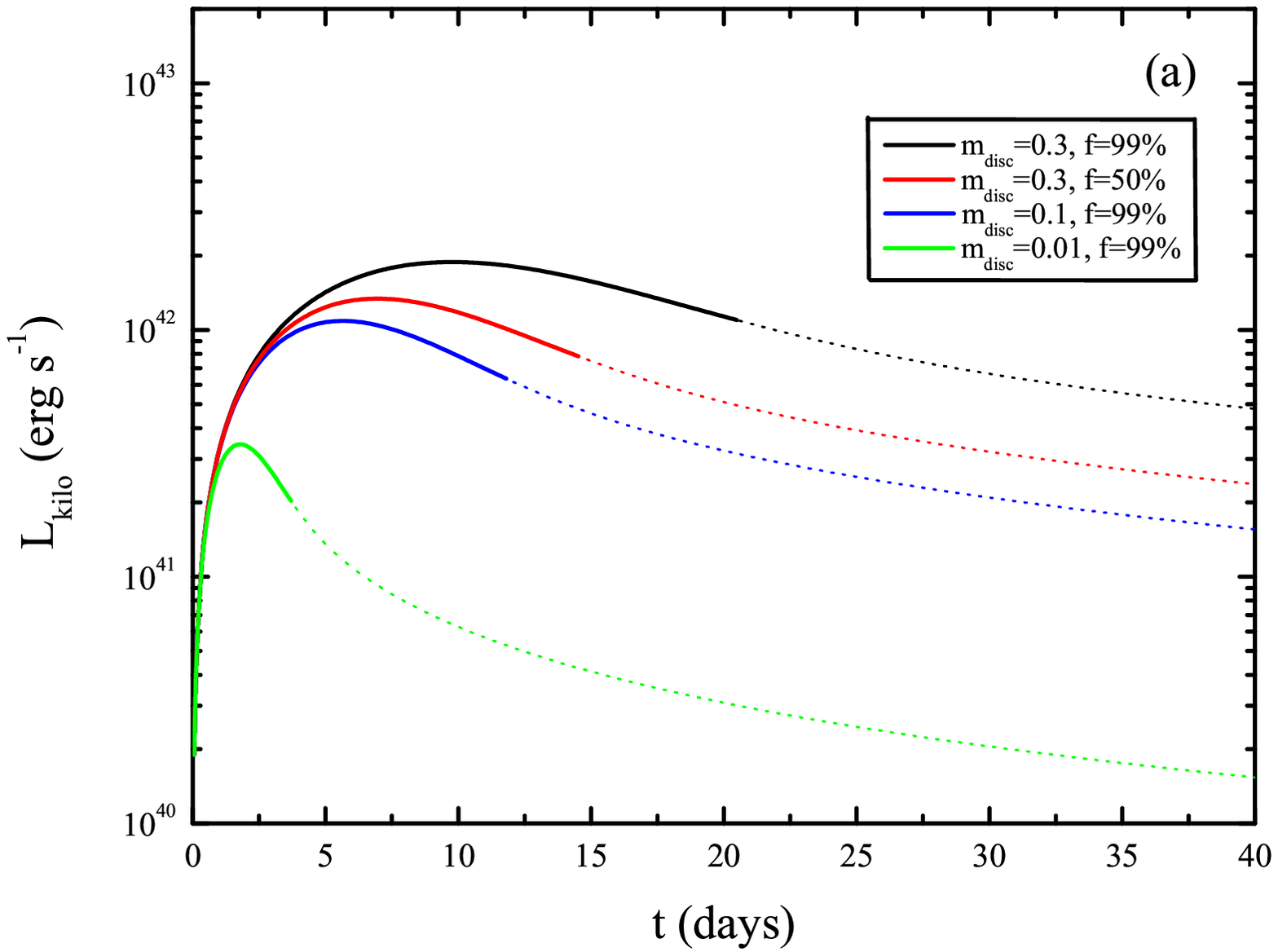}
\includegraphics[scale=0.5]{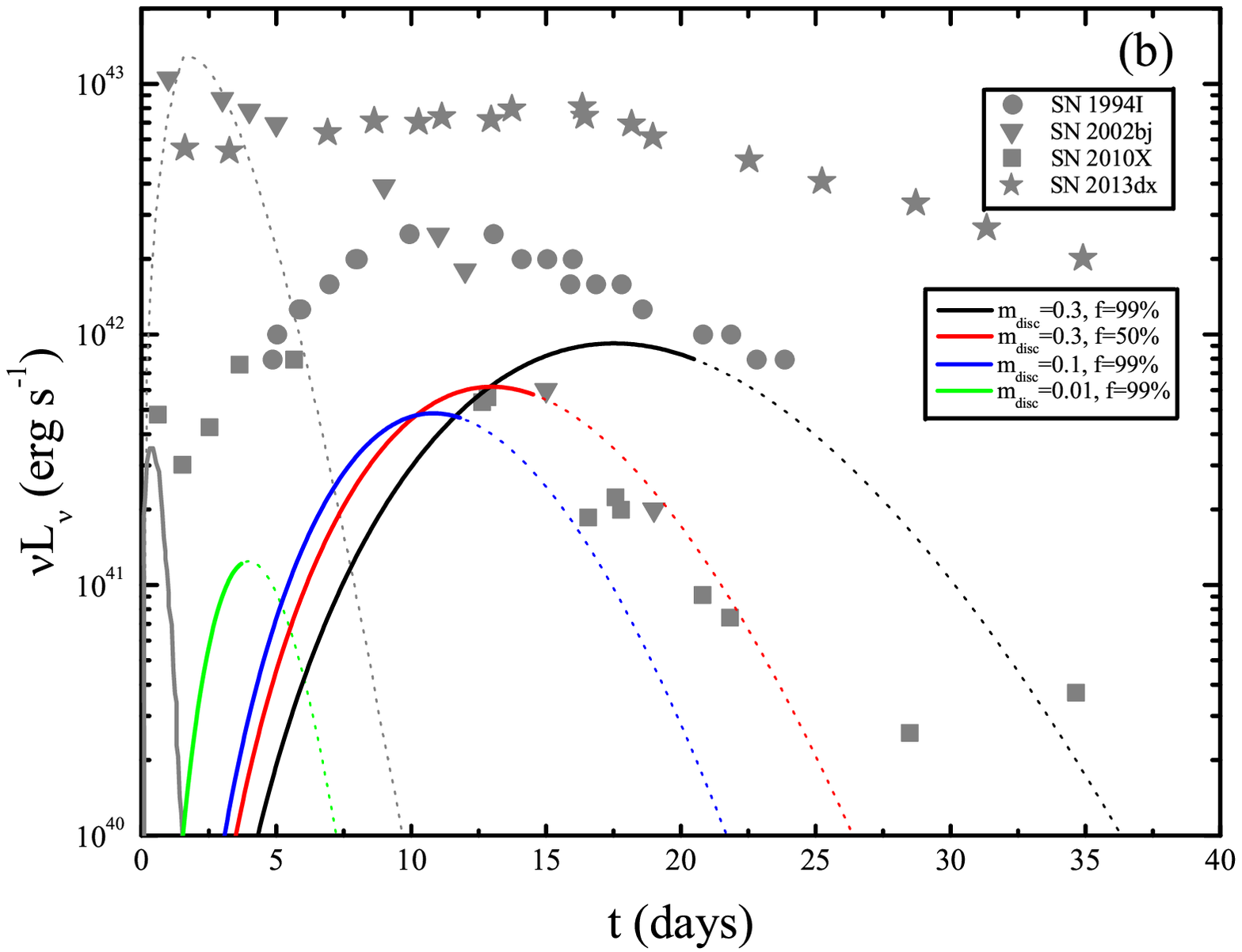}
\caption{(a) Total luminosity of kilonovae. (b) Comparisons of the optical ($\sim 1~\rm~eV$) light curves with SNe and mergernovae. Lines in different colors (black, red, blue, green) correspond to different values of $m_{\rm disc}$ and $f$. Thick solid parts and thin dotted parts indicate the expanding spheres in the optically-thick and optically-thin, respectively. The gray filled circles, triangles, squares, and stars represent SN 1994I, SN 2002bj, SN 2010X, and SN 2013dx data, respectively. The grey solid and dashed lines depict optical light curves of the millisecond-magnetar-powered mergernovae\citet[]{Yu2013}}
\end{figure}

The total luminosities of kilonovae are shown in Figure 2(a). Lines in different color (black, red, blue, green) correspond to different values of $m_{\rm disc}$, and $f$. Thick solid parts and thin dotted parts indicate the expanding spheres in the optically-thick and optically-thin, respectively. We notice that the luminosities of kilonovae increase with the increase of the outflow ratios and residual masses.

Figure 2(b) displays the optical ($\sim 1~\rm eV$) light curves of kilonovae, in comparison with SNe and mergernovae. The gray filled stars and circles respectively represent the data of SN 2013dx and SN 1994I. The grey solid and dashed lines depict the optical light curves of the millisecond-magnetar-powered mergernovae, which are adapted from Figure 3 in \citet[]{Yu2013}.

From Figure 1 and Figure 2, one can conclude that the energies of a GRB and its associated kilonova appear to be complementary to each other, mainly depending on the neutron-rich outflow ratio. Comparing those light curves, we find that for strong outflows and massive remnants, the durations of the kilonovae powered by the outflows are much longer than those of mergernovae, even approaching those of SNe. That is, the more massive accretion materials become the outflows, the more similar the behaviours of kilonovae become faint SNe, especially the SNe with the steep decay such as SN 2002bj and SN 2010X. Therefore, we prefer the name `quasi-SNe' for these phenomena, and we expect that a new type of `nova' like the faint SNe may be detected after merger events. In addition, the vertical distribution of the outflows might effect the luminosity of the kilonovae for different view angles.

\section{Summary}\label{Summary}

The progenitors of GRBs remain mysteries after about fifty years' discussions. It is still difficult to identify the physical origin of a GRB with multi-wavelength observational data available. The traditional definition of SGRBs and LGRBs by $T_{90}$ might not shed light on their progenitors.

After the mergers of NS-NS or BH-NS, a BH might may be born surrounded by a hyperaccretion disc. In the present work, we test the applicability of the BH hyperaccretion inflow-outflow model on powering both LGRBs and SGRBs in the compact binary merger scenario. If about half of the disc materials become outflows, the luminosity and duration of the hyperaccretion processes might satisfy the requirements of not only all SGRBs but also account for most of LGRBs. We also point out that, to verify the origin of GRBs one may need various information, including the characteristics of host galaxies, the SN associations, and the spectral lags, etc.

The optical/NIR emission observed in GRB afterglows are possibly powered by the numerous energy sources \citep[for reviews, see e.g.,][]{Metzger2017}. Here we propose a new mechanism of a BH hyperaccretion disc with extreme strong neutron-rich outflows, named quasi-SNe. The luminosities and timescales of quasi-SNe depend significantly on the outflow strengths. Consequently, there is a severe competition between GRBs and the associated quasi-SNe on the disc mass and energy budgets. In contrast, the luminosity of a mergernova depend on the energy injection from a magnetar, and there is no obvious correlation between it and the GRB luminosity, since most of the spin-down energy would be dissipated via GW emission \citep[e.g.,][]{Liu2017b}. In the particular case of GRB 170817A and AT 2017gfo, our model can naturally explain a weak GRB associated by a bright kilonova by considering the vertical distribution of the outflows, even without an off-axis jet for observers \citep[e.g.,][]{Lazzati2017,Xiao2017,Zhang2017}. We will investigate further this point in our future work.

Very possibly in the near future, more and more GWs' electromagnetic counterparts could be confirmed, produced by the compact binary mergers (especially for NS-NS and NS-BH). Noted that their optical/NIR emission might be originated from three mechanism: the ejecta or winds (kilonovae), the injection energy of magnetars (mergernovae), or the strong outflows from the hyperaccretion discs (quasi-SNe).

\section*{Acknowledgements}

We thank Zi-Gao Dai, Bing Zhang, and Tuan Yi for beneficial discussion and the anonymous referee for very useful suggestions and comments. This work was supported by the National Basic Research Program of China (973 Program) under grant 2014CB845800, the National Natural Science Foundation of China under grants 11473022 and U1431107, and the Fundamental Research Funds for the Central Universities (grant 20720160024).

\bsp
\label{lastpage}
\end{document}